# Growth of monolayer graphene on 8º off-axis 4H-SiC (000-1) substrates with application to quantum transport devices

N. Camara,[1,a)] B. Jouault,[2] A. Caboni,[1] B. Jabakhanji,[2] W. Desrat,[2] E. Pausas,[1] C. Consejo,[2] N. Mestres,[3] P. Godignon[1] and J. Camassel[2]

[1] CNM-IMB-CSIC – Campus UAB 08193 Bellaterra, Barcelona, Spain
[2] GES – UMR 5650 Université Montpellier 2/CNRS, 34095 Montpellier cedex 5, France
[3] ICMAB-CSIC, Campus UAB 08193 Bellaterra, Barcelona, Spain

Using high temperature annealing conditions with a graphite cap covering the C-face of an 8ºoff-axis 4H-SiC sample, large and homogeneous single epitaxial graphene layers have been grown. Raman spectroscopy shows evidence of the almost free-standing character of these monolayer graphene sheets, which was confirmed by magneto-transport measurements. We find a moderate p-type doping, high carrier mobility and half integer Quantum Hall effect typical of high quality graphene samples. This opens the way to a fully compatible integration of graphene with SiC devices on *the* wafers that constitute the standard in today's SiC industry.

It is now widely recognized that graphene-based devices are promising candidates to complement silicon in the future generations of microelectronic devices. To this end, the most favourable technique to produce graphene for industrial scale applications seems to be epitaxial graphene (EG) growth. This can be done by chemical vapour deposition on a metal,[1,2] or by heating a SiC wafer up to the graphitization temperature.[3-6] In the first case, the disadvantage is the need to transfer the graphene film on an insulating wafer. In the second case, the SiC wafer plays the role of the insulating substrate without any further manipulation. Of course, to be suitable for the microelectronics industry, these EG layers must be continuous and homogeneous at the full wafer scale or, at least, on surfaces large enough to process devices.

On the Si-face of the SiC substrates, graphitization at high temperature in an Ar atmosphere close to atmospheric pressure shows promising results for on-axis substrates. In this way, single layer epitaxial graphene (SLEG) and few layer epitaxial graphene (FLEG) have already been grown.[7,8] The main problem is that the growth of EG on the Si-face necessitates first the well known $6\sqrt{3}$ surface reconstruction. This reconstruction leads to a C-rich buffer monolayer on top of the SiC substrate. This buffer layer is insulating and the conduction take place only in the first "real" graphene layer on top of this buffer. In such case, this layer is not at all free-standing but strongly coupled to the C-rich buffer, heavily n-type doped and by consequence with a low carrier mobility. To some extent the problem can be solved either by application of an electric field through a high-k dielectric gate,[9,10] by thermal deposition of F4-TCNQ molecules,[11] or by hydrogenation of the buffer layer.[12] In such cases, the electrons density in the EG layer can be decreased and the mobility can reach values as high as 29 000 cm$^{-2}$/V·s.[11] Thanks to these treatments, the so-called "half integer" Quantum Hall Effect (QHE) - which is a clear evidence of homogeneous and high quality graphene films – has been observed several times on the Si-face of a SiC wafer.[9,11,13] However, the way to deal with this C-rich interface remains complex and the long term stability has not yet been demonstrated.

On the C-face of SiC substrates, the situation is entirely different. There is no longer any C-rich buffer layer at the interface and disordered FLEG exhibits similar band structure to graphene monolayers[14-16] with intrinsic mobility reaching ~ 30 000 cm$^2$/V·s.[17] Of course, this is only true for real turbostratic FLEG sheets – i.e. without AB stacking arrangements between adjacent layers- and when the film thickness remains reasonably uniform at the scale of the measuring device. These two requirements are hard to fulfil on commercial SiC wafers since the graphitization process on the C-face is mainly extrinsic and mostly starts at defects.[18] This makes the FLEG sheets not homogeneous with AB stacking arrangement recorded on such samples.[18] As a consequence, for a long time, no evidence of the "half integer" QHE typical for graphene layers has been found on these films. Recently, using a well controlled process, large SLEG areas have been produced on the C-face of on-axis SiC substrates and, on such monolayer, the QHE has been demonstrated.[19] The carriers were holes with mobility close to the one found in mechanically exfoliated graphene films on $SiO_2$/Si.[20] This proves the advantage and quality of SLEG grown on the C-face of a SiC wafer. For further integration of such graphene devices with current SiC technology (which usually includes the growth of several epitaxial SiC layers) on-axis substrates are not well suited. Off-axis substrates are preferred. Indeed, contrary to what happens with on-axis substrates, the homo-epitaxial growth of SiC layers on vicinal (8°off-axis) substrates prevents the formation of 3C-SiC inclusions and/or polycrystalline areas.[21] This is the reason why the 8°*off-axis* substrates constitutes the standard in modern SiC industry.

In a previous work, we have shown that there is an alternative route to grow EG on the C-face of a SiC substrate.[22] Under standard vacuum conditions, using a graphite cap to cover the sample, large single or bilayer graphene ribbons resulted. In the best conditions they could be ~ 600 μm long, having one or two monolayers of thickness. The main problem in this case was that the graphene layers grew mainly on large step-bunched terraces, leading to long (self-organized) graphene ribbons not wider than 5 μm. In this work, we use the same technology and show that much larger monolayers of graphene islands can be grown on semi-insulating 8°off axis 4H-SiC wafers. It has been demonstrated using scanning electron microscopy (SEM), atomic force microscopy (AFM), micro-Raman spectroscopy (μR) and Hall effect measurements.

To produce SLEG layers on 8°off axis SiC substrates, we use the recipes of Ref.22. with experimental results shown in Fig. 1. First, Fig. 1(a) shows a typical SEM picture of a SLEG island. It has a triangular shape and, similar to on-axis substrates, seems to nucleate from the initial defect shown in Fig. 1(b). This may be either an unintentional particle remaining on the surface, a crystallographic defect such as a threading dislocation or a simple scratch made by a diamond tip. Whatever the origin, the growth starts from one nucleating centre and expands in a 2-dimensions carpet-like way. All resulting triangles are self oriented, with the longest side following the (11-20) plane direction. After ~ 30 min graphitization, we obtained large monolayer graphene islands (~ 300 μm long and ~ 50 μm wide) shown in Fig. 1(c). In Fig. 2 we show the typical AFM image of a SLEG layer, with typical wrinkles showing evidence of the continuity and the strain-free character of these monolayers. Below the graphene islands, the step bunched areas of the SiC surface are clearly visible in both the SEM image of Fig. 1(b) and the AFM picture of Fig. 2. The corresponding terraces are typically 100 nm wide and 2 nm high and, similar to the work of Ref.22, the SiC surface covered by the SLEG layers differs from the uncovered one (on which no step bunching could be observed, neither by AFM nor by SEM).

Tens of similar monolayer islands have been probed by Raman spectroscopy using the 514 nm laser line of an Ar-ion laser for excitation. All spectra revealed that the islands are of the same nature and that even the largest ones (like the one shown in Fig. 1(c)) are homogeneous. The typical Raman spectrum of the low doped samples is displayed in Fig. 3, after subtracting the SiC substrate Raman signal for clarity. The D-band, which usually indicates the presence of disorder or edges defects, is very weak and the Raman signature is extremely close to the one found for exfoliated graphene on $SiO_2$/Si.[16] First, the 2D-band appears at low frequency (~ 2685 $cm^{-1}$) which is strong evidence that there is no strain at the layer to substrate interface (i.e. almost a free-standing SLEG layer). Second, this 2D-band can be fitted with a single Lorentzian shape with a FWHM of ~ 30 $cm^{-1}$.[23] Third, the ratio $I_{2D}$ / $I_G$ between the integrated intensities of the 2D-band and the G-band is high ~ 6, which suggests weak residual doping in the order of $10^{12} cm^{-2}$.[24]

Transport measurements have been performed, at low temperature on 10 different samples, using a maximum magnetic field of 13.5 T. The contact geometry allowed simultaneous measurement of both the longitudinal and transverse voltages, with the current flowing between two injection contacts at the flake extremities. From the sign of the Hall voltage we found that the carriers are holes –in agreement with the other results published on the C-face-[17,19] with concentration ranging from $1 \times 10^{12} cm^{-2}$ to $1 \times 10^{13}$ $cm^{-2}$ at low temperature, with weak temperature dependence. In the highly doped samples, Shubnikov-de Haas (SdH) oscillations were found. The plot of the inverse field, at which the oscillations maxima occurred, versus the Landau level index showed a clear linear dependence going down to the origin. This is the usual signature of the heavily doped graphene (see supplementary).

For the lowest doped layers, the transverse resistance also exhibited quantized Hall plateaus. They are clearly governed by the sequence $R_K / 4(n+1/2)$ with $R_K = h/e^2$ and $n = 0, 1, 2...$ This peculiar sequence of resistance values is the well-known quantum transport signature for the Dirac cone dispersion for monolayer graphene.[19] This is displayed in Fig. 4. In Fig. 4(a) we show the longitudinal and Hall resistance values for a low doped SLEG device with a hole concentration $n_s = 1.2 \times 10^{12} cm^{-2}$ and a mobility $\mu \sim 5000 \ cm^2/V \cdot s$ at $T = 1.6$ K. At $B = 12$ T the longitudinal resistance cancels, while the transverse resistance tends to 12.9 k$\Omega$ which is the expected value for the $n = 0$ plateau ($R_K / 2$). In Fig. 4(b) we present similar results, obtained on the same sample, after warming up to room temperature, standard cleaning with organic solvent and cooling down again to 1.6 K. The hole density, mobility and scattering processes have been significantly changed during this procedure. The new hole density is $n_s = 1.6 \times 10^{12} cm^{-2}$ and the mobility $\mu \sim 7\ 000\ cm^2/V \cdot s$. This makes the $n = 1$ plateau at $R_K/6$ much better resolved. This striking evolution of the electrical properties suggests that the mobility is highly sensitive to scattering processes extrinsic to the graphene layers, like adatoms, covalent bond impurities or PMMA residues which are probably also responsible of the p-type character of the layers. Finally, in Fig. 4(c) we show our best results obtained on the same chip with a similar device, with a lower doping $n_s = 8 \times 10^{11}\ cm^{-2}$ and a higher mobility $\mu \sim 11\ 000\ cm^2/V \cdot s$: the $n = 0$ plateau is now better resolved at low magnetic field and perfectly stable up to 13.5 T.

To summarize, we have shown the possibility to grow large islands of monolayer graphene on the C-face of 8°off-axis commercial 4H-SiC wafers. The graphene layers are continuous, almost free-standing and show quantum transport properties typical of high quality low doped SLEG film. Moreover, since the growth nuclei are usually polluting particles or defects, it should be possible to engineer the surface and grow SLEG selectively, at precise locations. Altogether, these findings open the way to a better integration of graphene with SiC microelectronic devices.

We acknowledge the EC for partial support through the RTN ManSiC Project, the French ANR for partial support through the Project Blanc GraphSiC and the Spanish Government for a grant Juan de la Cierva. N. C. also acknowledges A. Bachtold's, A. Barreiro and J. Moser from ICN Barcelona, for technical and theoretical support.

a) Corresponding author: camara.nico@gmail.com

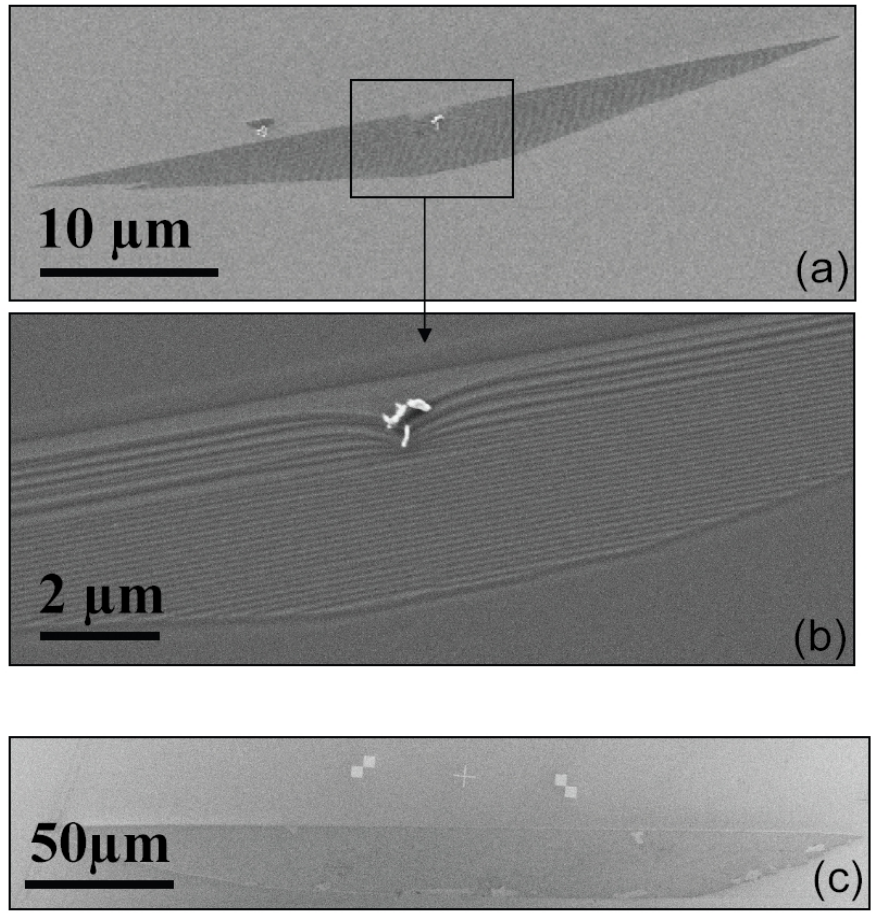

FIG. 1. (a,b) SEM images of a monolayer graphene island grown on the C-face of an 8°off-axis commercial 4H-SiC semi-insulating substrate. Both the starting nucleation point and the step bunching are clearly visible. (c) SEM image of one of the largest homogeneous SLEG islands grown in this work .

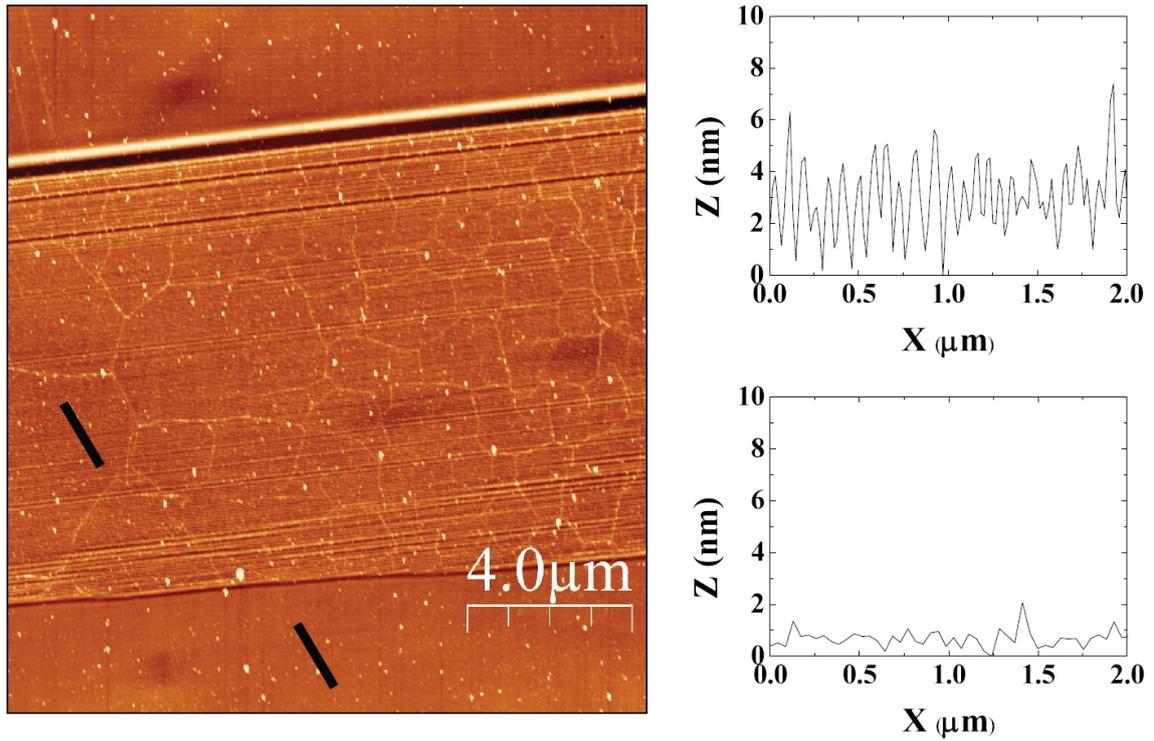

FIG. 2. AFM image of a SLEG layer grown on the C-face of an 8º off-axis commercial 4H-SiC semi-insulating substrate. Also shown are the typical SiC profiles collected inside the SLEG island (upper inset) where a clear step bunching is observed and outside (lower inset) where no surface reconstruction occurs.

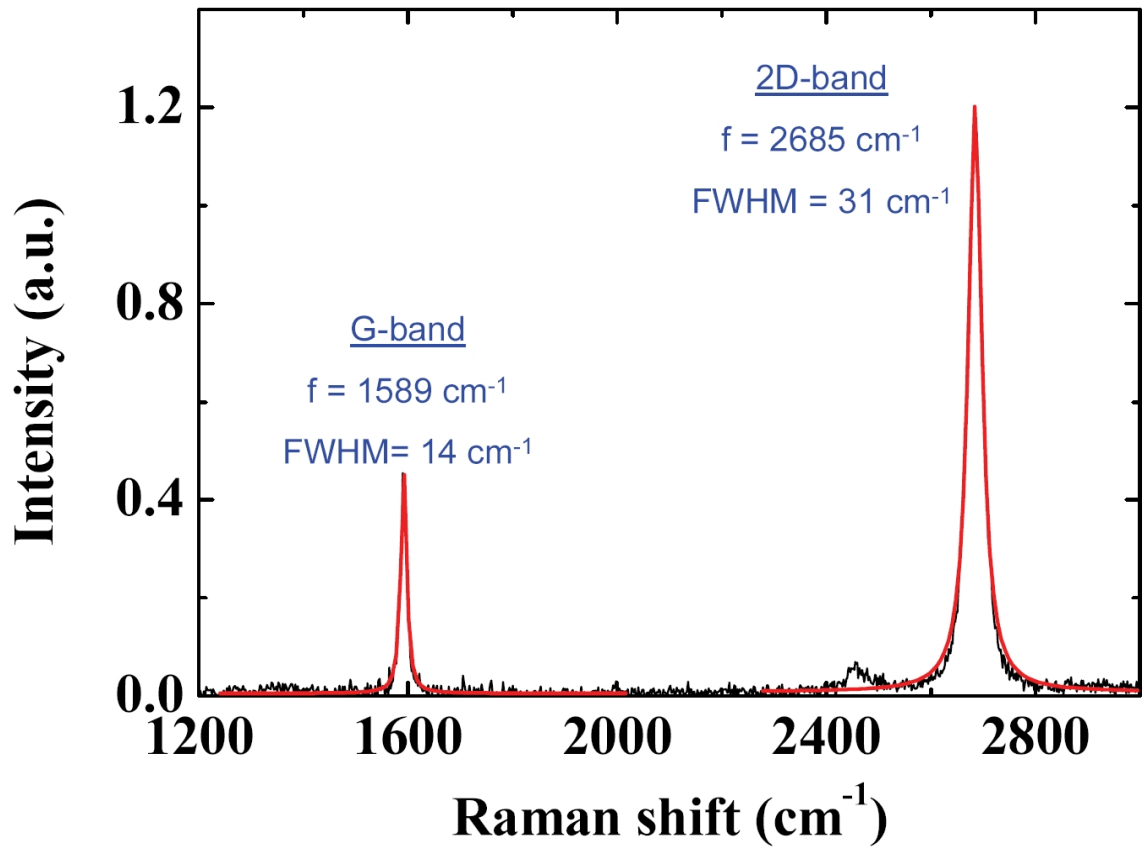

FIG. 3. Typical Raman signature of a SLEG film grown on the C-face of an 8°off-axis commercial 4H-SiC semi-insulating substrate. The red line displays the Lorentzian fit of the G- and 2D Raman modes. This spectrum has been obtained after subtraction of the SiC substrate Raman signal..

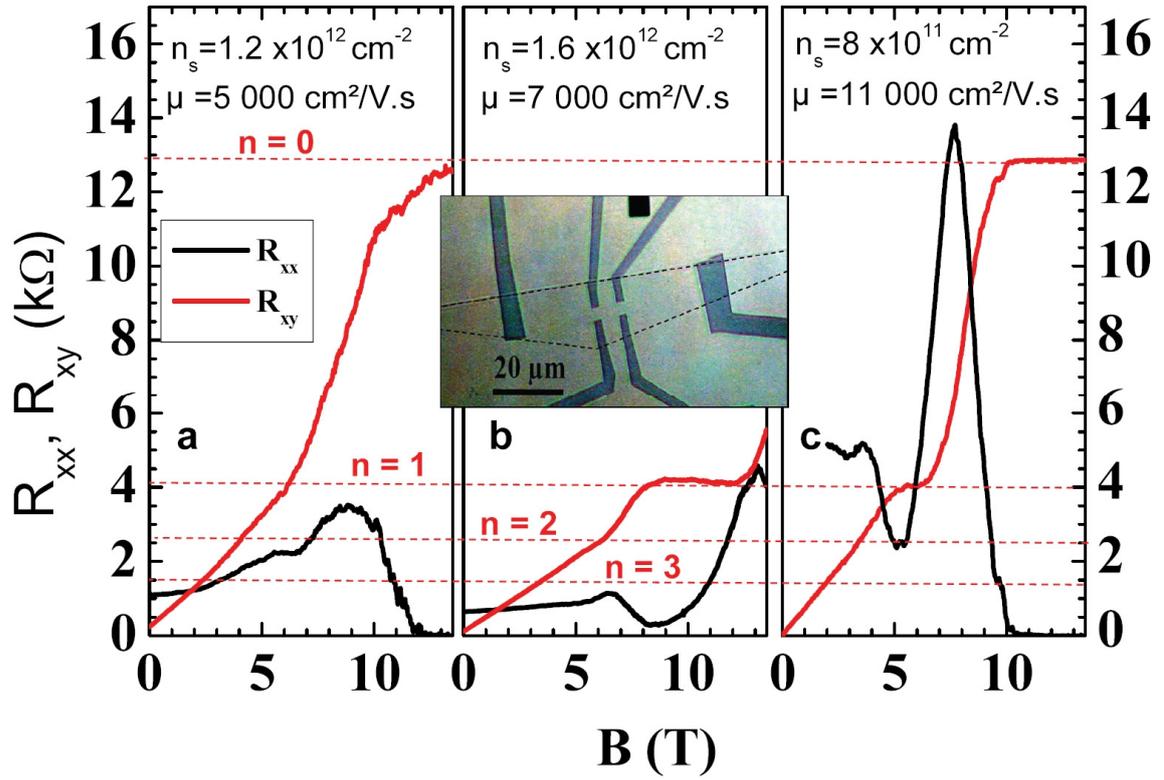

FIG. 4. Longitudinal and transverse resistance versus applied magnetic field B, at T = 1.6K. (a) The Hall resistance approaches the integer plateau $R_{xy}$ ~ 12.9 kΩ at B ~ 13T. The second plateau at 4 kΩ is hardly visible. (b) Similar measurements for the same sample, after warming up, cleaning and cooling down. The plateau n = 1 at 4 kΩ is better defined. In the inset is shown an optical microscopy image of the corresponding Hall bar. The total length of the contacted graphene ribbon is 50 μm, with a distance L = 5 μm between adjacent lateral probes and a lateral width of W = 10 μm. (c) Same measurements on the same chip for a similar device with a lower doping. The plateau n = 0 is now very well resolved while the other plateaus are hard to discern.